\documentclass{sig-alternate}

\usepackage{subfig}
\usepackage{array}
\usepackage{float}
\usepackage{url}
\usepackage{microtype}

\begin{document}

\title{Plausible Mobility: Inferring Movement from Contacts}

\numberofauthors{3}
\author{
\alignauthor
John Whitbeck \\
  \affaddr{Thales Communications and UPMC Paris Universitas}\\
  \email{john.whitbeck@lip6.fr}
\alignauthor
Marcelo Dias de Amorim \\
  \affaddr{CNRS and \\UPMC Paris Universitas}
  \email{marcelo.amorim@lip6.fr}
\alignauthor
Vania Conan \\
  \affaddr{Thales Communications}
  \email{vania.conan@fr.thalesgroup.com}
}

\conferenceinfo{MobiOpp}{'10, February 22-23, 2010, Pisa, Italy.}
\CopyrightYear{2010}
\crdata{978-1-60558-925-1/10/02}

\newcounter{copyrightbox}

\maketitle

\begin{abstract}
We address the difficult question of inferring plausible node mobility based only on information from wireless contact traces. Working with mobility information allows richer protocol simulations, particularly in dense networks, but requires complex set-ups to measure, whereas contact information is easier to measure but only allows for simplistic simulation models. In a contact trace a lot of node movement information is irretrievably lost so the original positions and velocities are in general out of reach. We propose a fast heuristic algorithm, inspired by dynamic force-based graph drawing, capable of inferring a plausible movement from any contact trace, and evaluate it on both synthetic and real-life contact traces. Our results reveal that (i) the quality of the inferred mobility is directly linked to the precision of the measured contact trace, and (ii) the simple addition of appropriate anticipation forces between nodes leads to an accurate inferred mobility.
\end{abstract}

\category{C.2.1}{Computer-Communications Networks}{Network Architecture and Design}[Store and forward networks]
\category{I.6.m}{Simulation and Modeling}{Miscellaneous}

\terms{Algorithms, Measurement}

\keywords{Delay Tolerant Networks, Graph Animation, Movement Inference}

\section{Introduction}
\label{sec:intro}

In the disruption-tolerant network (DTN) paradigm, mobile communication devices undergo a sequence of connections and disconnections from other devices forming \emph{contact opportunities}~\cite{dtn_fall_sigcomm}. Despite the growing interest in exploiting these contact opportunities for disseminating information under conditions when more traditional approaches are either impractical or unfeasible, there have been few real-life DTN deployments~\cite{Zhang:2007,kiosknet}. Instead, most evaluations of new protocols and designs have been done through simulations based on either synthetic mobility models or real-life contact traces. Neither is fully satisfactory. 

On the one hand, synthetic mobility models give full knowledge of the mobility and therefore allow for simulation of the specific features of radio channels (e.g., interferences and hidden stations) but do not accurately represent real-life (in particular human) mobility. On the other hand, contact traces are assumed to accurately represent real-life mobility but all geographical information is lost and simulators must make very simplistic assumptions on the communication channel (e.g., a node may only communicate with one of its current ``contacts'' at any given time~\cite{ONE}). A way out of this alternative could be the use of GPS measurements of human mobility~\cite{rhee:levy}.  Unfortunately, these are quite rare (only one such trace~\cite{rhee:levy} on CRAWDAD~\cite{crawdad}, as opposed to at least 5 Bluetooth contact traces). Furthermore, they are often unusable as contact opportunity information because the distances between devices are too great. Even if large scale GPS measurements~\cite{google_latitude} were able to achieve a density allowing one to consider geographic proximity as contact opportunities, they would still suffer from other limitations, such as not working indoors.

\textit{What if the information from the contact traces were sufficient to infer plausible node mobility?} The benefits would be twofold. First, being able to visualize node movement is in itself valuable, as it confers an intuitive understanding of the trace dynamics that can get lost in statistics. Second, using the inferred movement instead of simply contacts history would allow for a much finer simulation of the radio channel, particularly for dense contact traces~\cite{tournoux08_rollernet}, while retaining the realism captured by the contact traces. This paper makes the following contributions:

\begin{itemize}

    \item We define and discuss the problem of inferring plausible node mobility only from their contact information. To the best of our knowledge, this is the first time such a problem is addressed.

    \item We propose a formal definition of the problem as a system of non-linear inequalities.

    \item We describe and evaluate, both on synthetic and real-life contact traces, a heuristic but practical and effective method of inferring the movement of the nodes.

\end{itemize}

When only measuring the contact opportunities from an experiment with mobile devices, a lot of information is irretrievably lost. Consider a simple example with two nodes. When in contact, we can roughly locate them relatively to one another. However, when the time elapsed since the latest contact (inter-contact time) increases, the information regarding their relative distance decreases. After a while, it becomes difficult to say if they are still somewhat close or if they have gone in completely opposite directions. In a dense network, the higher the contact intensity, the more constrained our problem is. Although it is difficult to infer a mobility that is strongly correlated with the original mobility, we show in this paper that it is possible to propose a \emph{plausible mobility}, i.e., one that would have generated the same contact trace. This is examined in more detail in Section~\ref{sec:infering}. Since the ultimate goal is to improve simulations, inferring the exact mobility is not required. All we need is an inferred mobility that leads to better predictions.

In the next section, we position our paper in comparison with prior work. In Section~\ref{sec:infering}, we formally define the problem of inferring mobility from contact traces and discuss its challenges. In Sections~\ref{sec:heuristic} and~\ref{sec:evaluation}, we respectively propose and evaluate a heuristic approach to efficiently solve our problem. Finally we conclude our work and discuss the path ahead in Section~\ref{sec:conclusion}.

\section{Related Work}
\label{sec:related}

Delay/disruption-tolerant networks (DTN)~\cite{dtn_fall_sigcomm} arise when lack of end-to-end connectivity, rapidly changing topology, and/or potentially long communication delays render traditional mobile ad-hoc networks (MANET) approaches unfeasible~\cite{antonellis_classification}. Such networks encompass a vast spectrum of situations ranging from inter-planetary communications~\cite{dtn-interplanetary} to hop-by-hop data forwarding between portable devices to supplement an infrastructure for content dissemination~\cite{ioannidis09optimal}. In DTNs, node mobility can be exploited to increase the network capacity while compromising on delays by using a message \emph{store-and-forward} paradigm instead of the usual packet switching.~\cite{GrossglauserTse2002}.

Opportunistic mobile networks are a class of DTNs in which no knowledge of the future mobility of nodes is assumed. For example, this is the case of a network formed by the direct contact opportunities of hand-held devices, such as smartphones. Contact opportunities between mobile devices could be used either to replace or assist a wireless infrastructure for the dissemination of a given content. 
For example, Ioannidis et al. studied how to combine content pushing from a source in the infrastructure with opportunistic forwarding among subscribers in a way that ensures perceived content freshness from the subscribers while keeping the load on the infrastructure as small as possible~\cite{ioannidis09optimal}.

\begin{table}
  \centering
  \caption{Some wireless contact traces.}
  \includegraphics{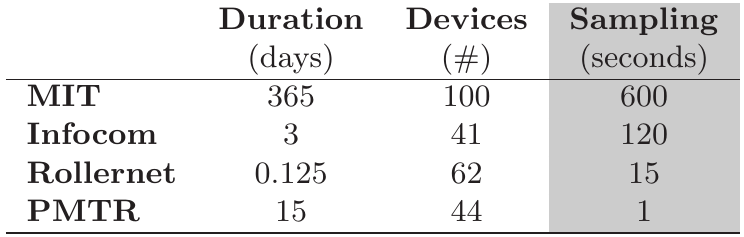}
  \label{table:traces}
\end{table}

A lot of effort has therefore gone into measuring human mobility, or at least the contact opportunities generated by human mobility. Direct measurements of human GPS coordinates have, to the best of our knowledge, only been performed by Rhee et al. in their work on human mobility models~\cite{rhee:levy}. These GPS measurements give accurate and fine-grained (10-second period) information but unfortunately only work when outdoors. Measurements of human contact opportunities overcome this indoors limitation but forgo location information. In the Reality Mining experiment conducted at MIT, each participation had a special application running on her/his mobile phone that captured proximity information from 100 subjects over an academic year~\cite{mit}. The Haggle project used Intel iMotes to capture the contacts between participants of the Infocom 2005 conference~\cite{chaintreau_mobility}. Both of these experiments relied on periodic Bluetooth scans. 

While Bluetooth has the advantage of being widely available, its scanning mechanism is too slow to effectively detect all contact opportunities. Indeed, the longer the sampling period (respectively 600 and 120 seconds for the MIT and Infocom traces), the more likely temporary link failures or short contacts will be missed. The Rollernet experiment, which also uses iMotes to capture the contacts in a rollerblading tour, was able to bring the sampling period down to 15 seconds~\cite{tournoux08_rollernet}. For finer measurements, a different beaconing method must be used. For example, Gaito et al. designed a specific device, a Pocket Mobility Trace Recorder (PMTR), and were able to measure contact opportunities every second~\cite{PMTR}, but the traces are not yet publicly available. Table~\ref{table:traces} compares these different traces. As we will see in Section~\ref{sec:heuristic}, shorter sampling periods translate into more constraints on our mobility inference problem, which in turn lead to solutions that better match the original mobility.

Inferring node mobility based solely on contact information is an open problem that has not yet received any attention in opportunistic networks. However, some similar questions have been addressed in other contexts. In wireless sensors networks, sensors can estimate their position relatively to a small number of \emph{anchor nodes} (typically equipped with a GPS receiver) using a variety of distance measurement techniques based on received signal strength or differences in beacon timings~\cite{wsn_localization}. 
In our approach, we must also rely solely on contact information and assume no low-level information on distances between nodes. Furthermore, unlike most wireless sensor networks, our nodes are all mobile. Finally, and perhaps most importantly, being free of decentralization or real-time requirements, our calculations take place offline and with full knowledge of future contacts.

\section{Inferring mobility}
\label{sec:infering}

\subsection{Problem definition}
\label{subsec:problem}

Let us consider a fixed number of mobile nodes. A contact trace is the list of contact events that occur between these nodes. Each event is recorded as a quadruplet consisting of the identity of both nodes, the instant when the contact is first established, and the instant when the contact goes down.

\vspace{2mm}\noindent\textbf{Real-life traces are noisy.} In real-life traces, depending on the scanning period and the choice of radio technology (e.g., Bluetooth, ZigBee), a number of contact opportunities may be missed, shortened, split, or merged. For example, using Bluetooth, neighborhood scans typically take several seconds and may not detect all reachable devices due to the frequency-hopping nature of the protocol. By using longer sampling periods, it becomes difficult to detect short contacts; even worse, a sequence of short contacts will likely be considered as a single long contact. Furthermore, a Bluetooth device may not simultaneously scan and respond to a scan. Therefore, many contacts will be missed simply because of the nature of the underlying protocol. Other wireless technologies, such as the custom-made Pocket Mobility Trace Recorders~\cite{PMTR}, can overcome these limitations but still have to contend with the traditional wireless issues such as interferences or hidden terminals. For these reasons, real-life traces must be considered noisy.

\vspace{2mm}\noindent\textbf{Synthetic contact traces.} Contact traces can also be extracted from synthetic mobility models by simulating a beaconing protocol or using a simple proximity-based model (i.e., a contact exists when two nodes are in transmission range of each other). Traces obtained in this fashion can be considered \emph{perfect}, in the sense that we have full control over all parameters and all contact opportunities are recorded. We will use this approach to evaluate our heuristic algorithm proposed in Section~\ref{sec:heuristic}.

\vspace{2mm}\noindent\textbf{Additional topology information.} Some nodes could have fixed and known positions, such as base stations in a 3G of Wi-Fi network, which enables us to place other nodes relatively to them. This approach has been well studied for node positioning in wireless sensor networks where some sensors have GPS capabilities~\cite{wsn_localization}. When all nodes are mobile, relative positioning information may still be available. For example, in the Rollernet experiment~\cite{tournoux08_rollernet}, an iMote was given to a member of staff that remained at front the rollerblading tour, while another was given to someone who stayed at the back. All other nodes in the trace must therefore be placed between these tail and head nodes. Finally, we could suppose that only the initial positions of the nodes are known. For synthetic traces, this information is readily available.

\vspace{2mm}\noindent\textbf{Plausible mobility.} Since we cannot hope to recover the exact initial mobility from a pure contact trace, we define the concept of \emph{plausible mobility}. In order to be \emph{plausible}, the inferred movement must (i) realistically limit the speed of the nodes and (ii) possibly produce the original contact trace, i.e. nodes in contact \emph{must} be within transmission range of each other while nodes not in contact \emph{should} be beyond transmission range.

In the end, our objective is to develop an algorithm that takes a contact trace and some additional information (e.g., fixed or relative positions) as input and generate a \emph{plausible} movement trace as output.

\begin{figure}[t]
  \centering
  \includegraphics{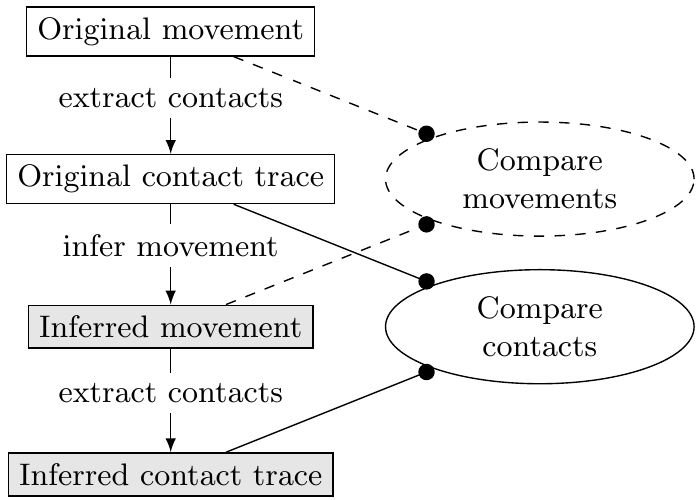}
  \caption{Evaluation framework.}
  \label{fig:evaluation}
\end{figure}

\subsection{Evaluation framework}
\label{subsec:framework}

When using synthetic mobility models, we initially have information of the nodes' mobility. From this, we can extract a contact trace, which we will use as input to our mobility inference algorithm. The inferred mobility can be compared to the original mobility, but can also be used to extract an inferred contact trace, which in turn can be compared to the original contact trace. When using real-life contact traces, we can no longer compare original and inferred mobility, but it still possible to compare the contact traces. This is summed up in Fig~\ref{fig:evaluation}.

We consider two ways of evaluating a mobility inference method: one comparing original and inferred mobility, and the other comparing original and inferred contact traces.

\subsection{Formalization}
\label{subsec:formal}

In this section, we describe what would constitute an ideal inference of mobility. The constraints defined below will guide us in the choice of the parameters for the heuristic approach proposed in Section~\ref{sec:heuristic}. Since the input is a contact trace, complete knowledge of past, present, and future contacts is assumed (offline inference).

\subsubsection{Definitions}
\label{subsubsec:definitions}

Let $N$ be the number of mobile nodes in the contact trace. These nodes move on a 2D plane, have a maximum speed $v_{\max}$ and a transmission range $r$. Let $(x_i(t),y_i(t))$ be the coordinates of node $i \in \{1,\cdots,N\}$. For the pair of nodes $(i,j)$, let $d_{ij}(t)$ denote the distance between $i$ and $j$ at time $t$. Furthermore, $T^\uparrow_{ij}(t)$ and $T^\downarrow_{ij}(t)$ denote, respectively, the time at which the next contact between $i$ and $j$ will appear and the time when the current contact will end.

At any time $t$ and any time interval $\Delta t$, the maximum node speed $v_{\max}$ imposes the following constraint on the positions of any node $i$:

\begin{equation}
\sqrt{{(x_i(t+\Delta t) - x_i(t))}^2 + {(y_i(t+\Delta t)-y_i(t))}^2} \le v_{\max} \Delta t.
\label{eqn:spd_cnstr}
\end{equation}

This constraint imposes that, given a valid solution at time $t$, and a short time interval $\Delta t$, the next valid position at time $t+\Delta t$ should be very similar.

\begin{figure}[t]
  \centering
  \includegraphics[scale=1.1]{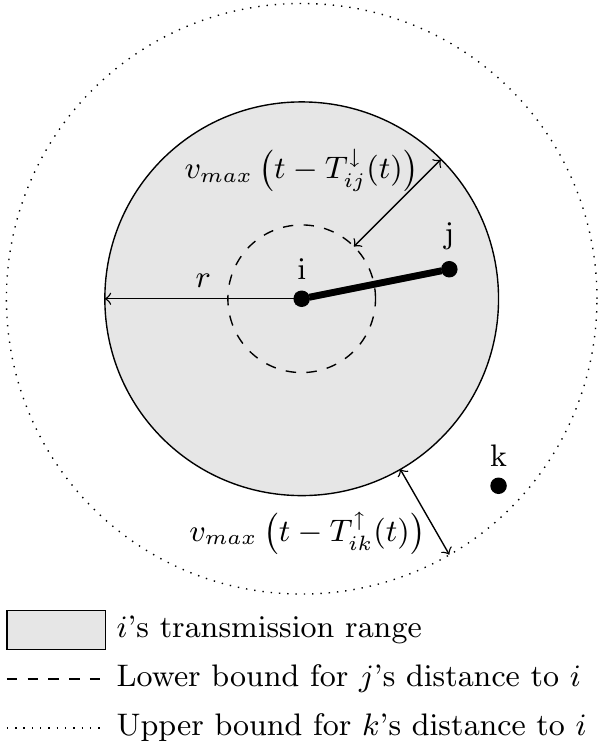}
  \caption{Constraints on the positions of nodes $j$ and $k$ relative to node $i$. Nodes $i$ and $j$ are in contact at time $t$ until $T^\downarrow_{ij}(t)$. A contact between nodes $i$ and $k$ will appear at time $T^\uparrow_{ik}(t)$. Parameters $r$ and $v_{\max}$ denote the transmission range and maximum speed, respectively. See Section~\ref{subsubsec:perfect}.}
  \label{fig:ranges}
\end{figure}

\subsubsection{Case 1: Synthetic contact traces}
\label{subsubsec:perfect}

In a synthetic contact trace, a contact appears when the distance between two nodes is less than $r$ and breaks when it is greater than $r$. Since we know when current contacts are going to break and when new ones will appear, we can further constrain nodes' positions. Indeed, nodes must get closer to each other before the contact appears and move away from each other before it goes down. This is illustrated in Fig.~\ref{fig:ranges}, where the contact at time $t$ between nodes $i$ and $j$ will go down at $T^\downarrow_{ij}(t)$ and a contact between $i$ and $k$ will appear at time $T^\uparrow_{ik}(t)$. As $t$ approaches $T^\downarrow_{ij}(t)$, node $j$ must get close to moving out of $i$'s transmission range. Relatively to $i$, $j$ must be able, given its maximal speed $v_{\max}$ to move out of transmission range at exactly $T^\downarrow_{ij}(t)$. Therefore, if $i$ and $j$ are in contact, the following constraint holds:

\begin{equation}
r - v_{\max}\left(t-T^\downarrow_{ij}(t)\right) \le d_{ij}(t) \le r.
\label{eqn:contact_cnstr}
\end{equation}

A similar analysis holds for the contact between $i$ and $k$. As $t$ approaches $T^\uparrow_{ik}(t)$, node $k$ must come closer to $i$'s transmission range. Relatively to $i$, $k$ must be able, given $v_{\max}$, to come within transmission range of $i$. Therefore, while $i$ and $k$ are coming into contact, the following constraint holds:

\begin{equation}
r \le d_{ij}(t) \le r+v_{\max}\left(t-T^\uparrow_{ik}(t)\right).
\label{eqn:not_contact_cnstr}
\end{equation}

\subsubsection{Case 2: Real contact traces}
\label{subsubsec:real}

While we know that a movement satisfying constraints~(\ref{eqn:spd_cnstr}), (\ref{eqn:contact_cnstr}), and~(\ref{eqn:not_contact_cnstr}) exists for the synthetic contact trace (i.e., the original synthetic movement), it is less clear for a real-life trace. Indeed, as we previously discussed, a real-life contact trace may be quite noisy and, in particular, miss many contacts. While this may seem like a simple relaxation of our constraints, it could in fact make the system unsolvable. Indeed, when considering real-life traces, the inclusive (i.e., \emph{in-contact}, Eq.~\ref{eqn:contact_cnstr}) and the exclusive (i.e., \emph{not-in-contact}, Eq.~\ref{eqn:not_contact_cnstr}) constraints no longer have the same importance. The inclusive constraint, based on the presence of a contact, can be trusted. However, the exclusive constraint, based on the absence of a contact, no longer strictly means that the distance between two nodes must be greater than the transmission range $r$. Indeed, one could imagine a node quickly passing by the other nodes, moving in and out their transmission ranges without triggering any contact detection. If we strictly enforce the exclusive constraint, such movements may no longer be possible.

\section{Heuristic solution}
\label{sec:heuristic}

In this section, we propose and evaluate a simple and efficient heuristic for inferring node mobility from their contact traces. Note that in order for it to have broad applicability, it should make as few assumptions as possible on the original mobility.

\subsection{Dynamic graph drawing}
\label{subsec:graph_drawing}

Our heuristic approach is inspired by works in \emph{dynamic graph animation}, even though its objective is quite different. Graph animation aims at (i) producing a sequence of readable and aesthetically pleasing representations of graphs and (ii) animating the transitions between successive graph layouts in a way that preserves the viewers' \emph{mental map} of the graph~\cite{onlineGD}. Sample applications include visualization of communication networks, social networks, and software library dependencies. Both goals are relevant to us. Not only do we wish to infer a sequence of positions for each node in the contact trace (i.e., a sequence of connectivity graph layouts), but real-life mobility intrinsically produces a sequence of connectivity graphs in which the transitions are easy to follow. However, while the function of dynamic graph animation is mostly aesthetic, our heuristic aims at meeting the constraints set out in Section~\ref{subsec:formal}.

In the context of graph animation, \emph{online} means that the graph layout algorithm is continuously running while new nodes or links appear and disappear on the fly, whereas \emph{offline} means that each successive graph is laid out separately. The offline method makes it difficult to preserve the viewer's \emph{mental map} during transitions, particularly long ones, between successive graph layouts. The online method procures an illusion of continuous mobility and allows for easy control of nodes' speed but is, in itself, insufficient in our case. Indeed, when a contact occurs between two nodes, they may, at that time, be very far away from each other in the online graph animation. This leads to a link in the connectivity graph that will, at least temporarily, straddle several connected components, which cannot constitute a satisfactory inferred movement.

Of particular interest to us are the \emph{force-based} layout algorithms~\cite{Fruchterman91graphdrawing}, in which attractive and repulsive forces are applied to nodes according to the connectivity graph. As in a real physical system, the nodes then converge to a minimum stress (or energy) position. Force-based algorithms are particularly well suited to our problem because each pair of nodes that are in contact will tend to be geographically close to each other.

Our heuristic for inferring a \emph{plausible mobility} from a contact trace will consist of running an \emph{online force-based dynamic graph layout} algorithm, built from the forces and refinements described in the next two sections.

\subsection{Forces}
\label{subsec:forces}

\begin{figure}
  \centering
  \includegraphics[scale=1.1]{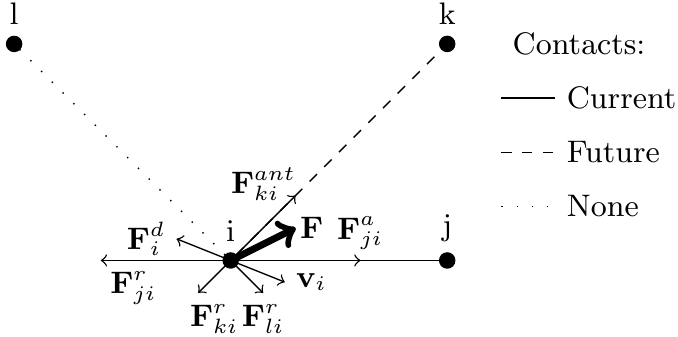}
  \caption{Forces applied to node $i$ from nearby nodes $j,k,l$.  $\mathbf{F}_{ji}^{a}$ is the attractive force of $j$ on $i$, $\mathbf{F}_{ji}^{r}$ is a repulsive force, and $\mathbf{F}_{ki}^{ant}$ is an anticipation force. $\mathbf{F}$ is the resulting force applied to $i$, bringing $i$ closer to $k$ before the $\{i,k\}$ contact appears. Full explanation in Section~\ref{subsec:forces}.}
  \label{fig:forces}
\end{figure}

As in a physical system, each node has a position, a speed, and an acceleration. All nodes have the same mass. Each node is subject to the four forces described below. The first three are classic, while the last is novel and necessary for the quality of our inference. We first present all of them before discussing how to set their parameters in the next section. Fig.~\ref{fig:forces} shows how these forces add up. Hereafter, $\mathbf{u}_{ij}$ is the unitary vector directed from $i$ to $j$.

\vspace{2mm}\noindent\textbf{Attraction.} Let $k$ be one of node $i$'s contact (i.e., a neighbor in the connectivity graph). $j$ attracts $i$ with a spring-like force:

\begin{equation}
\mathbf{F}^{a}_{ki} = K (d_{ik} - l_0 ) \mathbf{u}_{ik},
\label{eqn:attractive}
\end{equation}

\noindent where $d_{ik}$ is the distance between $i$ and $k$, $l_0$ is the spring's equilibrium position, and $K$ is a rigidity constant. This force contributes to keeping a pair of nodes in contact within transmission range of each other (right part of
constraint~(\ref{eqn:contact_cnstr})).

\vspace{2mm}\noindent\textbf{Repulsion.} Each node $j \neq i$ pushes node $i$ back with a coulomb-like force:

\begin{equation}
\mathbf{F}^{r}_{ji} =
\left\{
\begin{array}{ll}
- \frac{G}{(d_{ij}+\epsilon_0)^\alpha} \mathbf{u}_{ik} & \textrm{if } d_{ij} < d_{max}, \\
\mathbf{0} & \textrm{if } d_{ij} \ge d_{max},
\end{array}
\right .
\label{eqn:repulsive}
\end{equation}

\noindent where $G$ is an intensity constant, $\epsilon_0$ is small strictly positive constant to keep this force bounded, $d_{max}$ a cutoff distance beyond which this force no longer acts, and $\alpha$ a parameter that determines how this force's intensity decreases with distance. This force contributes to keeping  nodes that are not in contact away from each other (left part of constraint~(\ref{eqn:not_contact_cnstr})).

\vspace{2mm}\noindent\textbf{Drag.} In order to prevent excessive speeds (cf. constraint~(\ref{eqn:spd_cnstr})) and isolated nodes from moving too far away, each node $i$ is subject to a drag force:

\begin{equation}
\mathbf{F}^{d}_{i} = -D \mathbf{v}_i,
\label{eqn:drag}
\end{equation}

\noindent where $D$ determines how strong the drag is and $\mathbf{v}_i$ is the current speed of node $i$.

\vspace{2mm}\noindent\textbf{Anticipated attraction.} Since we have access to the entire contact trace, we know which nodes will meet node $i$ in the future. The idea here is to progressively make node $i$ move towards its future contacts, so that when the contact does appear in the trace, both nodes will be roughly in transmission range of each other (right part of constraint~(\ref{eqn:not_contact_cnstr})). If node $k$ meets $i$ at time $t_{ik}$, then, at time $t<t_{ik}$, $k$ attracts $i$ with a spring-like force:

\begin{equation}
\mathbf{F}^{ant}_{ki} = K (d_{ik} - l_0 ) e^{-\frac{(t-t_{ik})}{\tau}} \mathbf{u}_{ik},
\label{eqn:anticipated}
\end{equation}

\noindent which becomes equal, at $t=t_{il}$, to the regular repulsive force depicted in Eq.~\ref{eqn:repulsive}. The $\tau$ parameter characterizes when future contacts begin to have a noticeable influence on $i$'s movement.

\subsection{Issues and usage}
\label{subsec:usage}

When animating the graph, one must keep in mind that the goal is for each node to be within transmission range $r$ of its current contacts, and outside of range of all the other nodes. This should be encouraged but not strictly enforced, as it may otherwise lead to impossible configurations for real-life contact traces. Allowing the possibility of inexactitudes (i.e., adding or removing links in the inferred contact trace) is the price to pay for being able to infer movement from real-life traces.

A cluster of nodes can collectively have a strong repulsive force. As such, if another node comes into contact with one node in the cluster, they may never come into transmission range of each other, despite the attractive force. Setting a strong rigidity constant and setting the equilibrium length of the spring force to a point within the transmission range shown in Eq.~\ref{eqn:attractive} offsets this. In the rest of this paper, we use $K=30$ and $l_0=\frac{r}{2}$. In Eq.~\ref{eqn:repulsive}, we set $\alpha=\frac{3}{2}$ and $\epsilon_0=1$. This ensures that the repulsive force (i) is bounded by $G$ and (ii) does not decrease too quickly. Finally, we choose $G$ in Eq.~\ref{eqn:repulsive} so that the distance between two nodes in contact, absent all other forces, converges towards $\frac{3}{4} r$.

An issue not usually addressed in the graph drawing community is how to deal with disconnected components. Since we are handling DTN contact traces, we cannot avoid this problem, as the connectivity graph is almost always split into several disconnected components and many isolated nodes. Freivalds et al. propose laying out each connected component separately and then using a \textit{packing} algorithm to place them relatively to one another~\cite{graph_packing}. However, this completely ignores that, in our case, the relative placement of connected components should not be arbitrary. Fortunately, our \emph{anticipated attraction} force circumvents this problem by creating attractive forces between disconnected components and thus guiding their relative movement, orientation, splits, and merges. The value of $\tau$ in Eq.~\ref{eqn:anticipated} is a tradeoff. Small values of $\tau$ mean that only the very short term future is considered for animating the contact trace, while large values can create so many constraints that no movement is possible. Good values are linked to the characteristic evolution time of the connectivity graph. Finally, cutting off the attractive force eliminates long range interactions between nodes that could interfere with the initially weak \emph{anticipated attraction} forces. In the rest of this paper, we use $d_{max}=3r$.

\section{Evaluation}
\label{sec:evaluation}

In this section, the heuristic described above is applied both on a synthetic mobility model (Random Waypoint - RWP) where the contact trace is considered perfect and on the real-life contact trace with the shortest available sampling period, Rollernet~\cite{tournoux08_rollernet}. Note that our heuristic makes absolutely no assumptions about the underlying mobility model. In fact, it is particularly poorly adapted to Random Waypoint, since in it the nodes try to avoid each other, whereas in RWP nodes pay no attention to each other. Nevertheless, we still manage to infer \emph{plausible} movements.

\subsection{Synthetic movement}
\label{subsec:synthetic}

\begin{figure*}[t]
   \centering
   \includegraphics[scale=0.9]{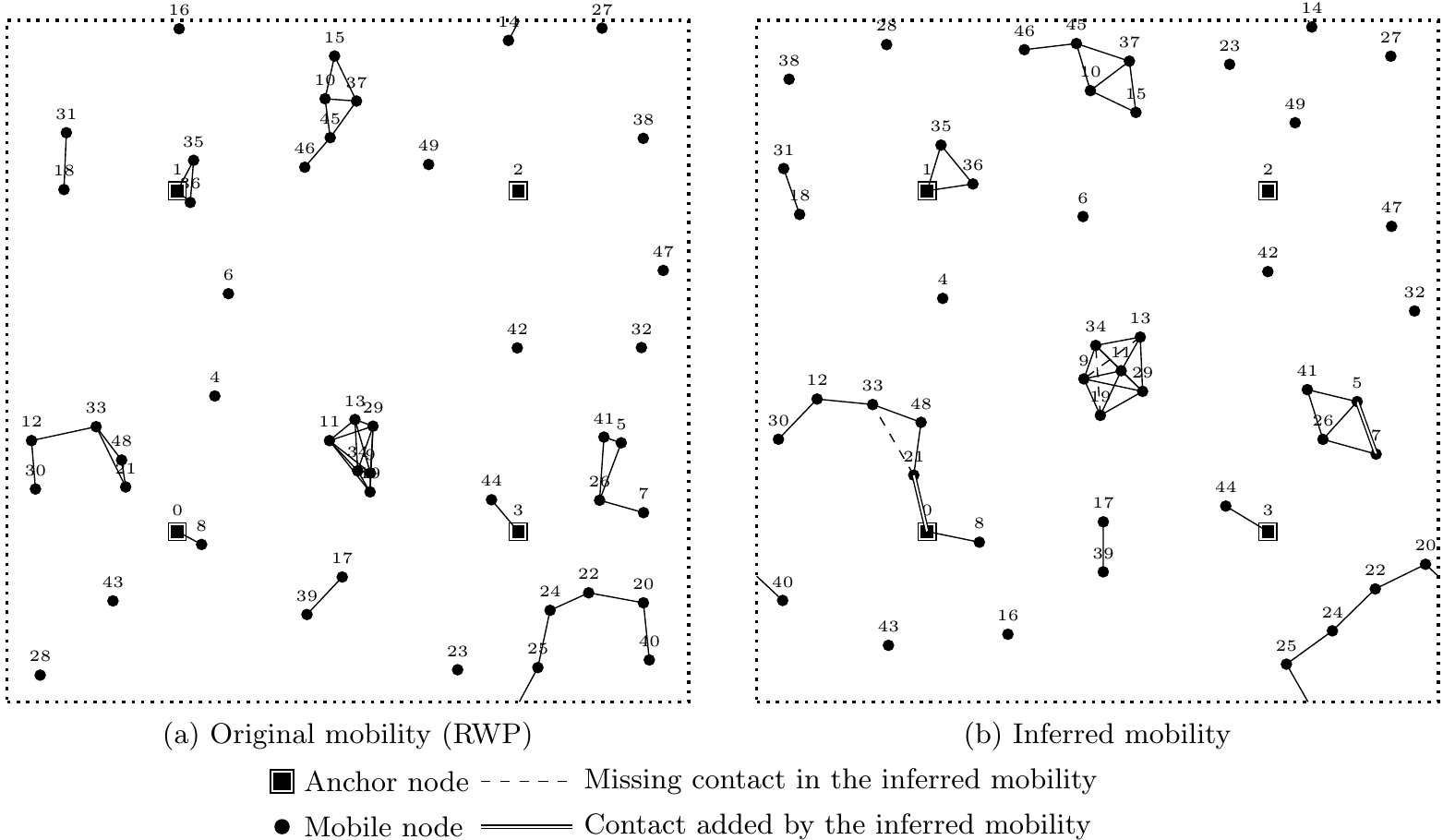}
   \caption{Mobility snapshots of the original Random Waypoint movement (a) and the movement inferred from its contact trace (b) after 200 seconds. Nodes 0, 1, 2, and 3 are fixed \emph{anchor nodes}. The other 46 nodes move on a torus.}
   \label{fig:example}
\end{figure*}

The synthetic mobility scenario considered in this section consists of 50 nodes moving according to the RWP model on a 1000m x 1000m torus, so as to avoid any border effects. RWP has some well known shortcomings, such as a usually non-uniform steady state spatial distribution of nodes and gradual speed decay~\cite{rwp_harmful}. However these are irrelevant in this work since the steady state spatial distribution on a torus is in fact uniform, and we only run short RWP simulations (300 seconds) aimed solely at producing random contact traces with which to study our mobility inference algorithm. Furthermore, we consider that 4 nodes, hereafter called \emph{anchor nodes}, are immobile and symmetrically distributed on the torus at positions (250m,250m), (250m,750m), (750m,250m), and (750m,750m). These anchor nodes serve as reference points for placing all the other nodes. In a real-world scenario they could wireless access points or 3G base stations. Unless otherwise noted, nodes move at speeds chosen uniformly between 1m/s and 10m/s with no pause time, and their transmission range is 100m. For each run of this mobility scenario, a contact trace was extracted from successive snapshots of the nodes' positions with a certain sampling period, by simply considering that any pair of nodes within transmission range of one another are in contact. A snapshot of this mobility is seen in Fig.~\ref{fig:example}a.

Through some experimentation, the $\tau$ parameter of the anticipated attraction force was set to 5. This means that future contacts start significantly pushing their nodes closer to each other 5 seconds before the contact actually appears. Smaller values meant that contacts scheduled to appear in the original trace would only show up later in the inferred trace, because of the delay until both nodes would get into transmission range. Greater values meant that each node would be attracted to a larger subset of the other nodes. If too big, this can result in preventing most movement. Additionally, the initial positions of all nodes were known to our heuristic.

\begin{figure}[t]
  \centering
  \subfloat[0 fixed nodes (0.64)\label{fig:dist_corr_f0}]{\includegraphics{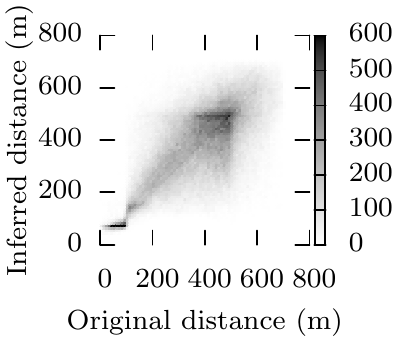}} \quad
  \subfloat[4 fixed nodes (0.82)\label{fig:dist_corr_f4}]{\includegraphics{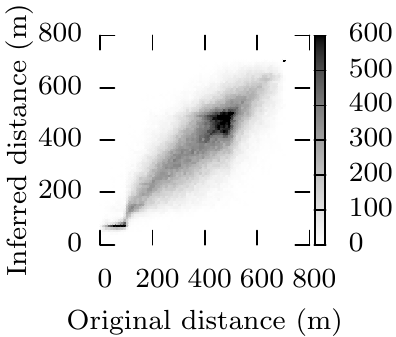}}
  \caption{Correlation between pairwise distances in a synthetic mobility trace and its inferred mobility. Correlation values are given in parentheses.}
  \label{fig:pairwise_dist_corr}
\end{figure}

We first compare the inferred mobility to the original. While we do not expect to be able to infer the exact node positions, the relative distances between pairs of nodes should be correlated. Indeed, Fig.~\ref{fig:pairwise_dist_corr} examines the correlation between the pairwise distances in the original and inferred mobility. Every time step, the distance between each pair of nodes was measured in both the original and inferred mobility. Fig.~\ref{fig:dist_corr_f0} represents the correlation scatter plot in the absence of anchor nodes, while in Fig.~\ref{fig:dist_corr_f4} the four anchor nodes are present. The presence of a few fixed anchor nodes clearly helps positioning the other nodes. 

One artifact of our heuristic is visible for small distances. Indeed, all distances below the transmission range (100m) in the original mobility are roughly mapped to 75m in the inferred mobility. This is a result of the equilibrium distance of two nodes in contact as discussed in Section~\ref{subsec:usage}. An example is visible on Fig.~\ref{fig:example} where nodes 1, 35, and 36 are closely grouped in the original mobility (Fig.~\ref{fig:example}a) but appear as a triangle in the inferred mobility (Fig.~\ref{fig:example}b).

\begin{figure}[t]
	\centering
	\includegraphics{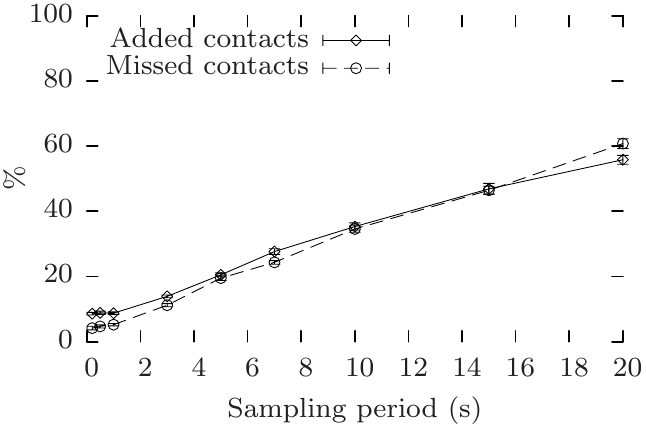}
	\caption{Influence of contact sampling period on quality of mobility inference. Values expressed as percentage of existing contacts.}
	\label{fig:sampling}
\end{figure}

\begin{figure}[t]
	\centering
	\includegraphics{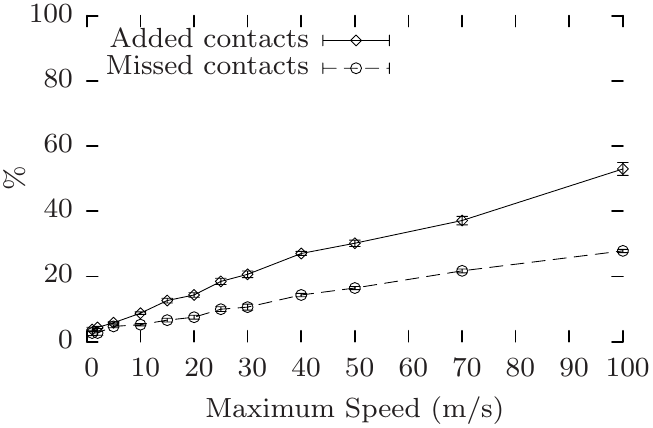}
	\caption{Influence of maximum node speed on quality of mobility inference. Values expressed as percentage of existing contacts.}
	\label{fig:mobility}
\end{figure}

We then compare the inferred contact trace to the original. Every time step, we check the constraints~(\ref{eqn:contact_cnstr}) and~(\ref{eqn:not_contact_cnstr}). If two nodes that should be in contact are not within transmission range of each other, that contact is considered \emph{missed}. Conversely if two nodes not in contact are within transmission range of each other, an \emph{added} contact is counted. In Figs.~\ref{fig:sampling} and~\ref{fig:mobility}, \emph{missed} and \emph{added} contacts are expressed as a percentage of the number of existing contacts at that time.

This is a rather strict way of comparing two contact traces, as even slight delays in the start or end of a contact will register as respectively a missed or added contact for that time period. Fig.~\ref{fig:sampling} shows the effect of the sampling period of the original contact trace on the quality of our inference. The proportion of both added and missed contacts increases with the sampling period. This is due to several reasons. Firstly, as the sampling period increases, it becomes more difficult to assume that a contact present in one period but not the next lasted the whole period. Lacking other information, our heuristic does however make this assumption. During a time period, a given node, when pushed by the anticipated attraction force towards its next contact, is still restrained by the attractive forces of the nodes that were in contact with it at the beginning of the period. For longer sampling periods, we may overestimate the duration of many contacts and therefore prevent a node from moving towards its future contacts. This translates into both \emph{missed} contacts, from not getting within transmission range of new contacts in time, and \emph{added} contacts from remaining within transmission range of old contacts. Secondly, smaller sampling periods also catches short contacts that would otherwise be ignored. These provide many extra contacts that a node must meet on the way to meeting its next contacts according to the longer sampling period, and thus enable a much smoother and progressive mobility inference. 

High-frequency contact measurements are therefore necessary in order to achieve good mobility inference. This likely holds for any real or synthetic mobility, and any inference algorithm. Unfortunately, no such high-frequency contact traces are publicly available.

Fig.~\ref{fig:mobility} examines how the dynamic of the underlying mobility impacts the quality of the inference. The sampling period here is 1 second, and we use different values for the maximum speed in the Random Waypoint model. Both added and missed contacts increase with the maximum node speed. The more dynamic the movement, the harder the inference. Again, we expect this to hold for any combination of mobility models and inference algorithms.

\subsection{Real-life traces}
\label{subsec:reallife}

\begin{figure}
   \centering
   \includegraphics{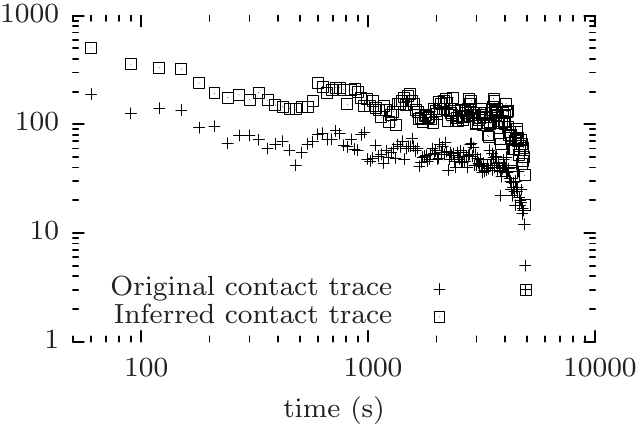}
   \caption{Inter-contact time distribution for original and inferred mobility in Rollernet~\cite{tournoux08_rollernet}.}
   \label{fig:rollernet_ict}
\end{figure}

As discussed previously, the longer the sampling period in a contact trace, the more difficult it becomes to claim that a contact translates into a link that lasts roughly as long as the sampling period. Several contact traces (see Table~\ref{table:traces}) were considered, but in this section we focus on the Rollernet trace~\cite{tournoux08_rollernet}, collected during a rollerblading tour around Paris, which due to its low sampling period (15 seconds), is the one that comes closest to capturing the evolution of the connectivity graph.

Compared to the synthetic mobility in the previous section, this contact trace results from a real-life experiment and thus must be considered lossy. Furthermore, the nodes are highly mobile and the average contact duration (26 seconds) is barely longer than the sampling period (15 seconds). A look at Fig.~\ref{fig:sampling}, may lead one to assume that a 15-second sampling period is too long for a satisfactory mobility inference. However, unlike in the previous section, these sampling periods are not synchronized across devices. The contact trace thus provides us with a continuous, though incomplete, stream of contact events to guide our heuristic.

When inferring mobility for the Rollernet contact trace, we used the same parameters as in the previous section. There are no fixed anchor nodes in the trace, but the head and tail nodes of the rollerblading tour are known. No participant skated ahead (resp. behind) the head (resp. tail) node. The head and tail nodes were constrained to moving along a horizontal axis. If ever a node wants to pass the head node for example, then the head node is moved to ensure that it remains ahead. This enables all nodes to be placed relatively to the head and tail nodes. Due to the rapid contact process in the Rollernet trace, the anticipated attraction forces are sufficient to keep the rollerblading tour compact and naturally lead to the emergence of the accordion phenomenon~\cite{tournoux08_rollernet} because the head and tail node get closer when the contact density increases and pushed apart when it decreases. The resulting mobility is aesthetically pleasing and helps guide intuition when working on the dataset. 

Fig.~\ref{fig:rollernet_ict} plots the inter-contact time distribution for the first 5000 seconds of both the original and inferred Rollernet contact traces. As usual, both distributions follow a truncated power law~\cite{chaintreau_mobility}. However, the inferred contact trace contains many more contacts than the experimental one. Indeed, on average, it adds about 36\% extra contacts and misses about 9\% of the measured ones. This may seem like a lot, but unlike in the previous section, some of those added contacts may in fact be real contact opportunities that the Bluetooth devices simply failed to pick up. Are we introducing bogus contact opportunities? Or are we identifying contacts that the experiment missed? Or both? The lack of reliable high-frequency contact traces make it impossible to decide.

\section{Conclusion and further work}
\label{sec:conclusion}
In this paper, we examined a new and interesting problem, the inference of a \emph{plausible} mobility from a wireless contact trace. Indeed, mobility is more difficult to measure but enables better simulations, particularly in dense networks, whereas contact information is easier to measure but only allows for simplistic simulation models. Our heuristic solution, based on ideas from dynamic graph drawing, can animate any wireless contact trace, while making practically no assumptions on the mobility that produced the contact trace. Our results highlight the need for reliable high-frequency contact traces in order to extract a \emph{plausible} mobility from a wireless contact trace.

This work, a first approach of a difficult problem, can be pursued in several directions. A more comprehensive study of our proposed heuristic must be undertaken using different mobility models and scenarios. In particular, the time to reach a threshold of accuracy when starting \emph{without} knowledge of initial positions in the experiments of Section.~\ref{subsec:synthetic} should be examined. Furthermore, other heuristics, perhaps closer to the constraints in Section~\ref{subsec:formal}, should be explored. Finally, the idea that infering mobility from a contact trace before running simulations on it ought to lead to more realistic results than just using the contact trace alone, must be properly tested. For example, one could compare the performance of various network protocols in an opportunistic network based on simulations, with complete knowledge of the mobility, with only contact information, and with the \emph{plausible} mobility infered from the the contacts. The results on the infered mobility should be closer to those using the real mobility than to those using only the contact information.

\section{Acknowledgements}
\label{sec:acknowledgments}
This work has been partially supported by the ANR project Crowd under
contract ANR-08-VERS-006.

\bibliographystyle{abbrv}
\bibliography{whitbeck_mobiopp10}

\end{document}